\documentclass[12pt,preprint]{aastex} 
\usepackage{apjfonts}

\newcommand{\pf}{$\rightarrow$} 
\newcommand{\kms}{km\,s$^{-1}$}

\newcommand{\vlsr}{v$_{\rm lsr}$}

\newcommand{\cmmm}{cm$^{-3}$}

\newcommand{\Mo}{M$_{\odot}$}

\received{2003 February 5}
\accepted{2003 August 12}
       
\begin{document}       
\title{NGC 2264 IRS1: the central engine and its cavity}

\author{K. Schreyer} 
        \affil{Astrophysikalisches Institut und Universit\"atssternwarte
               (AIU) Jena, Schillerg\"a{\ss}chen 2--3, D--07745 
               Jena, Germany}
        \email{martin@astro.uni--jena.de} 
\author{B. Stecklum and H. Linz}   
        \affil{Th\"uringer Landessternwarte Tautenburg, Sternwarte 5, 
        D--07778 Tautenburg, Germany}
\and \author{Th. Henning} 
        \affil{Max--Planck--Institut f\"ur 
   Astronomie (MPIA), K\"onigstuhl 17, D--69117 Heidelberg, Germany}

\begin{abstract} 
We present a high--resolution study of NGC 2264 IRS1 in CS 2$\rightarrow$1 
and in the 3-mm continuum using the 
IRAM Plateau de Bure Interferometer. We 
complement these radio data with images taken at 2.2 $\mu$m, 4.6 $\mu$m, 
and 11.9 $\mu$m. The combined information allow a new interpretation of the 
closest environment of NGC 2264 IRS1. 
No disk around the B--type star IRS1 was found. IRS1 and its low--mass 
companions are located in a low--density cavity which is surrounded by 
the remaining dense cloud core which has a clumpy shell--like structure.
Strong evidence for induced on--going star formation was found in 
the surroundings of IRS1. A deeply embedded very young stellar object  
20$''$ to the north of IRS1 is powering a highly collimated bipolar 
outflow. 
The object 8 in the closer environment of IRS1 is a binary 
surrounded by  
dusty circumbinary material 
and powering two bipolar outflows.
\end{abstract}
%

%
                      \section{Introduction} 
%

The near--infrared source NGC 2264 IRS1 (IRAS 06384+0932, Allen 1972)
is a relatively isolated luminous infrared source (distance 800 pc) in 
the star--forming region NGC 2264. Previous observations 
\citep{schwartz,Krugel,Phillips,Schreyer97} 
suggest that IRS1 is a young star located within a dense molecular core. 
This core is embedded in a more extended CO envelope. The 2--200 $\mu$m 
luminosity of $\approx$ 3500 L$_{\odot}$ 
(Harvey,  Campbell, \& Hoffman, 1977)
implies an early B--type zero--age main--sequence star \citep{schwartz}.

Previous ground--based near--infrared images  
(Scarrott \& Warren--Smith 1989; Schreyer et al.\ 1997) 
show a twisted  jet--like feature to the north of NGC 2264 IRS1, which 
could be a gas stream ``piercing'' the surrounding dark cloud. Two 
outflow systems were reported in this region 
\citep{Schreyer97}, 
one at the position of IRS1 and a second one associated with a deeply 
embedded small star cluster $\approx 47''$ to the southeast of IRS1. 
Ward--Thompson et al.\ (2000)
found a ridge (NGC 2264 MMS) of bright submm- and mm--emission around IRS1 
wherein five emission peaks (MMS\,1--5) are located. MMS\,3 is associated 
with the gas clump at the position of the small star cluster 
\citep{Schreyer97}  
MMS\,4 and 5 are located on the eastern and southern side of IRS1,
respectively.
Wang et al.\ (2002) 
found a number of bipolar jets in the region of NGC 2264 MM which point 
to complex star formation processes in the whole area (diameter $\le 2'$).
%
%
%
The high resolution H$^{13}$CO$^+$ 1\pf0 map of Nakano, Sugitani, and Morita (2003) 
shows a shell of dense gas clumps around IRS1. Their corresponding continuum map 
displays four more compact millimeter sources, three in the closer surroundings 
of IRS1, and one is associated with the small star cluster to the southeast 
(MMS3).


In this paper\footnote{Partly
based on observations collected at the European Southern Observatory, 
La Silla, Chile, under programme IDs 62.I-0530 and 66.C-0209.
}$^,$\footnote{Based on observations carried out with the IRAM 
       Plateau de Bure Interferometer. IRAM is
   supported by INSU/CNRS (France), MPG (Germany) and IGN (Spain).}, 
we focus on  the infrared source IRS1 and its nearest 
environment (diameter $\le 51''$). We studied this region in 
the $K$(2.2$\mu$m)--band with ESO's NTT as well as in the CS 2\pf 1 line 
and the 3-mm continuum using the Plateau de Bure Interferometer (PdBI).
Additional thermal infrared data came from observations with 
ESO's mid-infrared camera TIMMI2.
The aim of this study was to answer the following questions:
What is the reason for the displaced centre of the bipolar outflow 
seen with single--dish telescopes (e.g.\ 
Schreyer et al.\ 1997)?
Does the presence of a ``wiggled jet'' possibly originating from IRS1 
imply the existence of a disk around IRS1\,? 
Is this twisted jet dense enough to be traced with the PdBI 
in order to obtain some information about the velocity structure? 
Is there an opposite ``jet--like'' feature\, to confirm the 
suggestion of a (bipolar) jet? Finally, we wanted to learn what 
the nature of the central engine is.

%
%
                     \section{Observations}
%
%

Near--infrared images of NGC~2264~IRS~1 were obtained with SOFI at
ESO's New Technology Telescope (NTT) on 1999 March 3 in the Ks band.
These observations were performed in polarization mode, i.e., with the
Wollaston prism and a slit mask at a pixel scale of 0\farcs295. After
standard image processing (flat--fielding, sky removal and bad--pixel
correction) the total intensity image resulted from the co--addition of
four frames taken at polarizer orientations of 45$^\circ\!\!,$ 
90$^\circ\!\!,$ 135$^\circ\!\!,$ and 180$^\circ\!\!.$
The total integration time amounts to 384 seconds. The observing conditions
were very good, leading to a FWHM of the stellar images of 0\farcs57.
Furthermore, for the central overlap region of the stripes (covering 
IRS1 and its near vicinity) these data were used to deduce a polarisation map.

Imaging at 11.9$\mu$m (pixel scale 0\farcs 202) and 4.6 $\mu$m (pixel
scale 0\farcs 315) was performed on 2001 March 18 with TIMMI2 
\citep{Reimann}
at ESO's 3.6-m telescope. Chopping was done perpendicular 
to nodding; the offset throw for both movements was 20$''$ at 11.9$\mu$m 
and 30$''$ at 4.6 $\mu$m, respectively. We restored the original field of 
view by applying the Projected Landweber restoration method 
(Bertero, Boccacci, \&  Robberto 2000) 
to the data, which we modified for our purposes 
(see Linz et al.\ 2003).

With the PdBI 
\citep{Gui}, 
we observed CS 2\pf1 as well as the corresponding continuum emission  
at 97.98 GHz. Successful observations were obtained with five 15-m antennas  
in August and November 1998 using baseline lengths between 21--254 m. 
We applied one correlator unit 
with a total bandwidth of 10 MHz centred at the CS    2\pf1  
line (\vlsr\ = +8.0 \kms) which leads to a velocity resolution 
of 0.12 km s$^{-1}$ ($\widehat{=}$ 0.039 MHz). 
Two spectral correlator units, each with a bandwidth  
of 160 MHz, were assigned to the measurements of the continuum. 
The band pass and phase  
calibration was performed with the objects 
3C454.3,  0528+134, and 0748+126.   
%
Compelled by a flux loss of $\approx$50--80\% in the PdBI CS 2\pf1 data
compared to our previous IRAM single-dish data (Schreyer et al. 1997),
we combined our IRAM 30-m CS 2\pf1 map (extent = 2$' \times$ 2$'$ 
= twice as large as the PdB primary beam) and the PdB interferometer 
data in order to fill up information about the missing flux between 
that zero-spacing and the shortest spacing from the interferometer. 
The IRAM map was Fourier--transformed and then fiddled into the
PdB visibility data. Hereby, we benefit from the fact, that the
diameter of the single-dish telescope (30~m) is twice as large as the 
one of the interferometer telescopes (15~m), and that we have also 
included short baselines (down to 21~m) for the interferometer 
measurements. Thus, regarding the density of uv points, there is a 
relatively smooth transition from the contribution of the single-dish 
data to the contribution of short--baseline interferometer
data. With the visibilities resulting from the Fourier--transformed 
single--dish map we completely cover the uv points of the shortest
PdB baseline.
Maps of 256$\times$256 square pixels with 0.5$''$ pixel size  
were produced by Fourier--transforming the calibrated  
visibilities, using natural weighting.  
The synthesized beam sizes  (HPBW) are  
3.05$''$$\times$1.82$''$  
for the continuum data  
and 3.21$''$$\times$1.95$''$ for the line map  
(with zero--spacing correction), 
each with a position angle of 21$^{\circ}\!\!.$  
For the data reduction and the final phase calibration, we used   
the Grenoble Software environment GAG.

%
%
             \section{Results and data analysis} 
             \subsection{Infrared results} 
%
%

\begin{figure}[t]
\vspace{15cm} 
\includegraphics{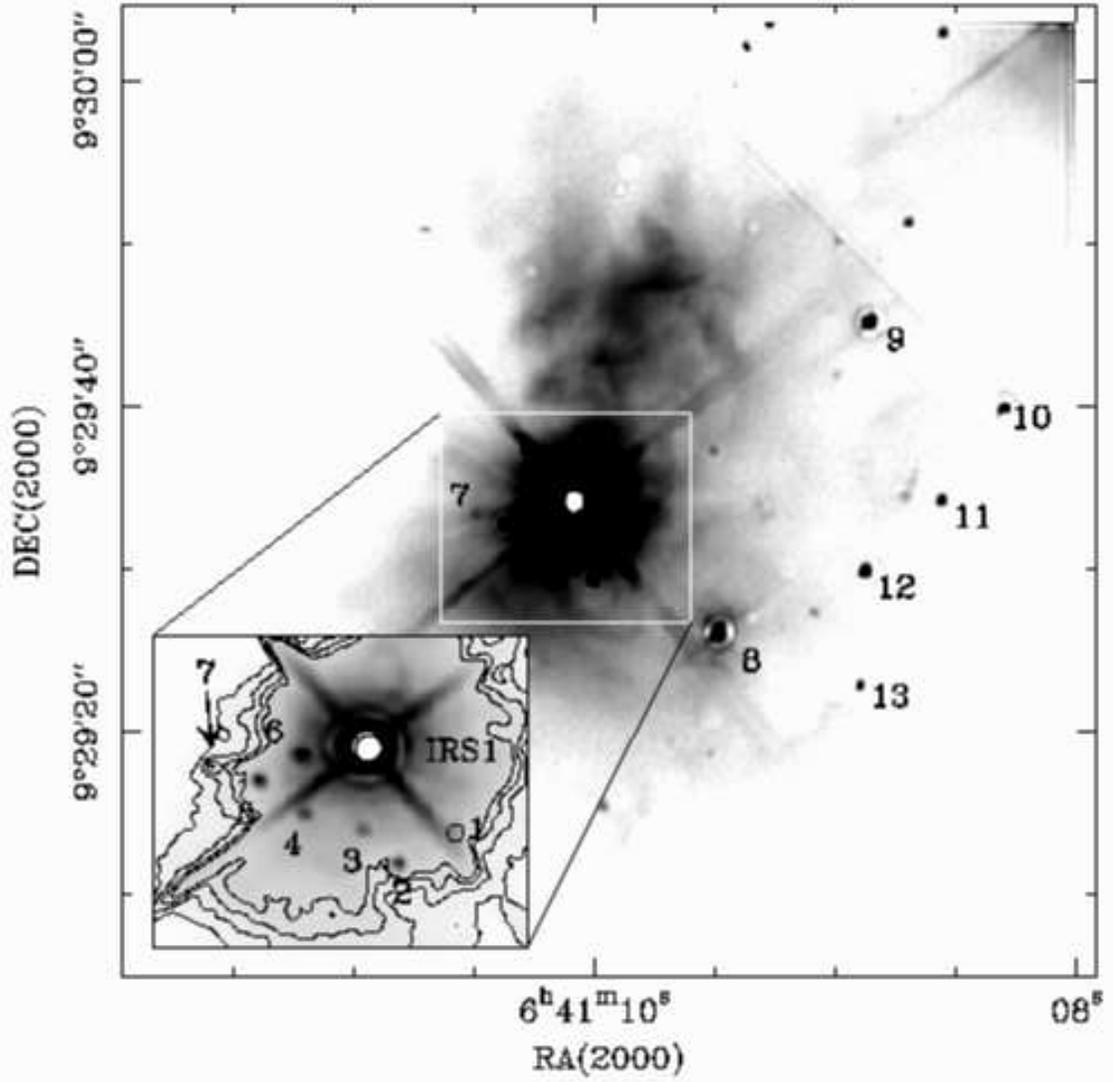} 
\figcaption{ \small
        The NTT $K$--band image, whereby the 
        large image is shown with very low cuts to emphasize 
        fine details, e.g., the zig--zag structure to the northwest. 
        The contour lines are 0.11\%, 0.15\%, 0.19\%, and 0.23\% of
        the peak value.
\label{ntt}   
}  
\end{figure}

\begin{figure}[t]
\vspace{15cm} 
\includegraphics{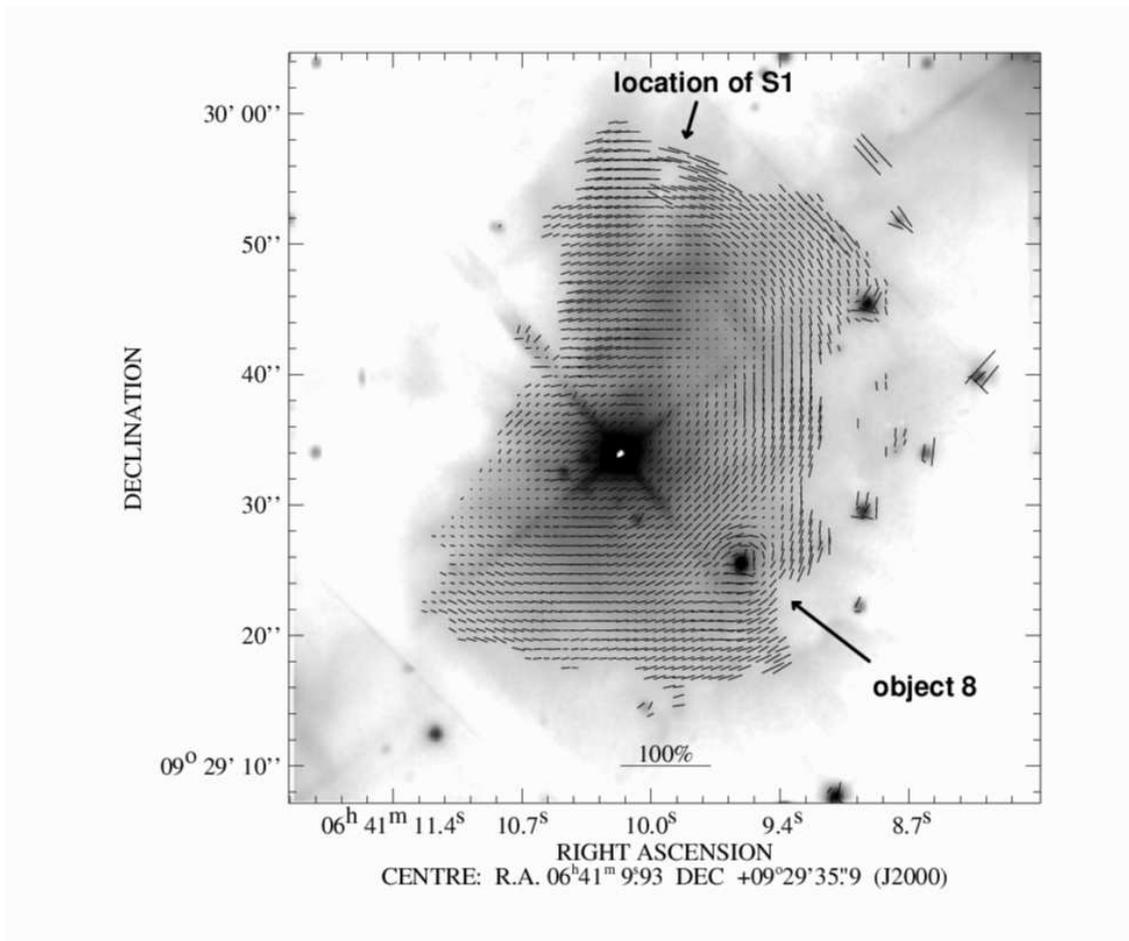} 
\figcaption{ \small
The $K$--band polarisation map in the region around IRS1.
Object 8 shows its own centrosymmetric polarization pattern, confirming 
that this object is surrounded by circumstellar
dust. 
Note the decrease in intensity at the location of the mm--continuum source 
S1 (see Fig.~\ref{csfig}) 
which also causes a localised disruption of the polarisation pattern.        
\label{pol}   
}  
\end{figure}

\begin{figure}[t]
\vspace{18cm} 
\includegraphics{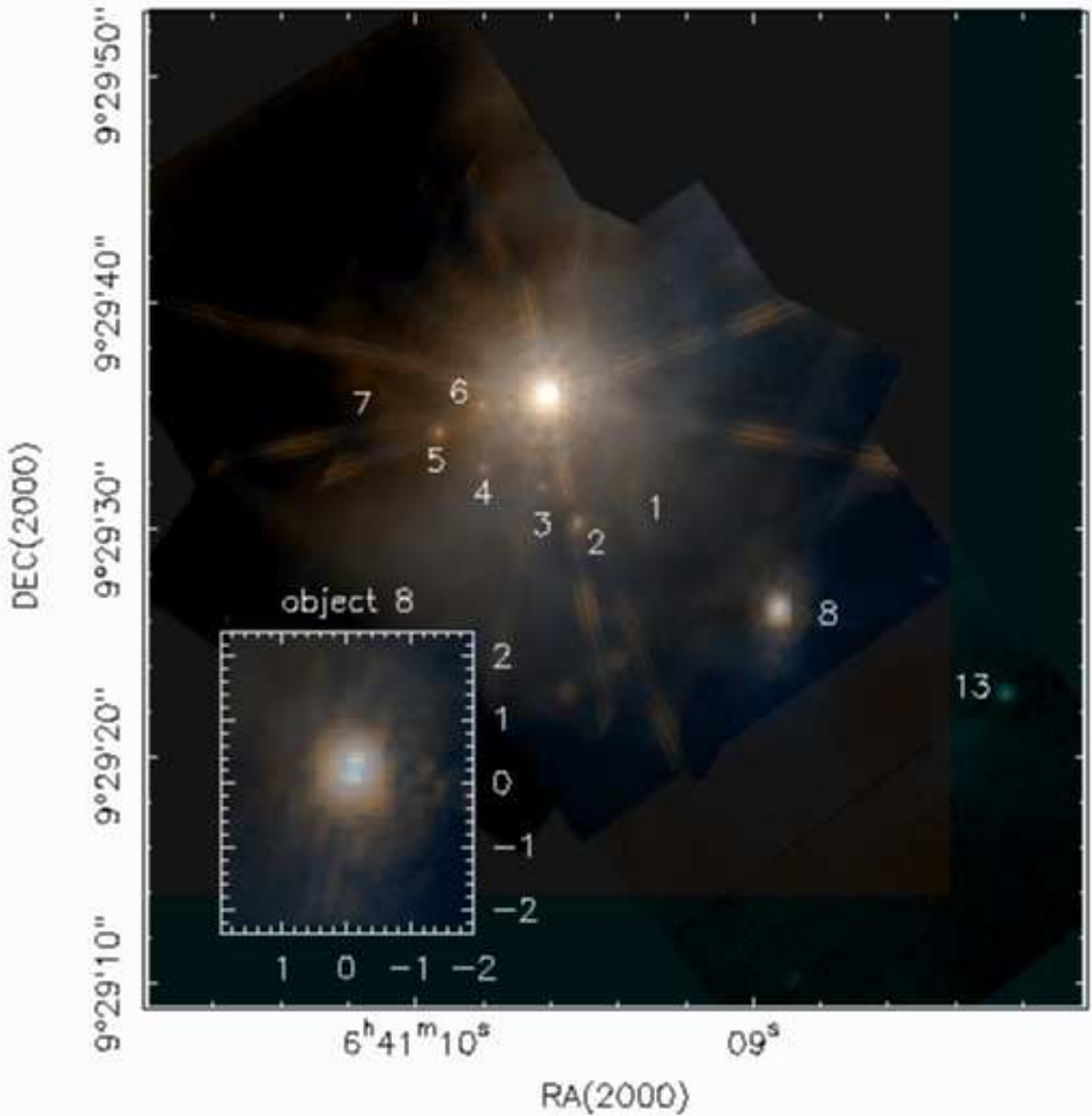} 
\figcaption{ \small
        The NICMOS mosaic image is composed of the following three images:
        the 1.6-micron 
        image (blue), the mean of 1.6-micron and 2.2-micron images 
        (green), and the 2.2-micron image (red).
\label{nic}   
}  
\end{figure}

Fig.~\ref{ntt}  
and Fig.~\ref{pol} show 
the NTT $K$--band image 
and the $K$--band polarisation map, respectively. 
%
The six low--mass companions to the south 
of IRS1 coincide with the 
%
faint
%
objects shown in the HST NICMOS image obtained by 
Thompson et al.\ (1998)
who labeled these 
%
as
%
object 1 to 6. An additional 
very faint object 7 was found 
%
 (see Table~\ref{pos}) 
%
which was not recognized by Thompson et al.\ 
since it was close to the diffraction spikes 
in the NICMOS images. The other most 
%
intense
%
$K$--band stars, seen in the NTT image, are denoted with 8 to 13.
Object 8 seems to be of special interest since it is not
point--like but marginally extended in 
%
the
%
NS direction. While the overall polarization is caused by 
scattered light from IRS~1, object 8 shows its own 
centrosymmetric polarization pattern
%
(see Fig.~\ref{pol}), 
%
confirming 
that this object is surrounded by circumstellar
dust. Thus, we identify object 8 as an young stellar object in the
immediate vicinity of IRS~1. This result is confirmed by archival 
NICMOS images taken for testing the coronograph (Proposal ID. 7808). 
Fig.~\ref{nic} is a composite we constructed from these NICMOS images 
which shows that object 8 is a binary. 
%
%
The projected 
separation of the two components amounts to 0.27$''$$\pm$0.01$''$ 
(= 216 AU) which is somewhat larger compared to the GG Tau binary 
system with a projected separation of 38 AU.
In addition, the faint object 7 is present in this figure.

The NTT image exhibits a similar jet--like feature -- as found in the 
previous low--resolution images by 
Hodapp (1994) and  Schreyer et al. (1997) 
-- which looks like a twisted gas stream to the north of IRS1. 
However, both, the NTT and the NICMOS images do not 
clarify the true origin of this feature.
We note that, based on the images alone,  
the illuminated features to the north could be also material 
which ablates from the surface 
of a denser cloud clump behind IRS1.

\begin{figure}[t] 
\vspace{18cm} 
\includegraphics{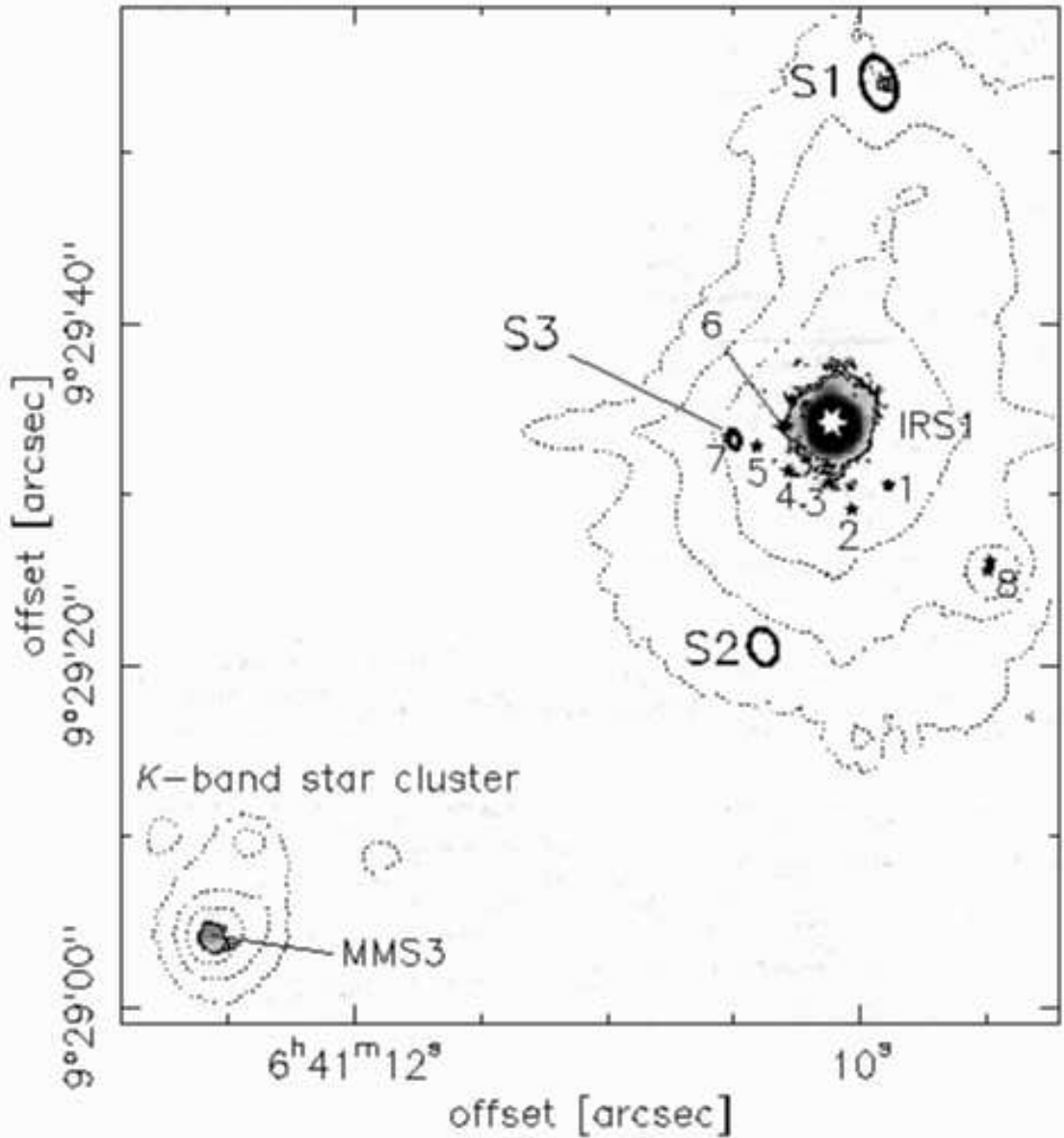}
\figcaption{ \small
        The TIMMI2 image at 11.9 $\mu$m (grey-scale image) 
        with 6$\sigma$ and 8$\sigma$ (thin black) contour lines 
        is overlaid with the thick 50\% contours (ellipses) of 
        the 3mm-continumm point sources S1, S2 and S3 (see 
        Sect.~\ref{3mm}), and the dotted contours of the $K$-band 
        image of Schreyer et al.\ (1997). The three 11.9 $\mu$m sources 
        correspond to IRS1, S1, and MMS3.
\label{timmi2}   
}  
\end{figure}

With TIMMI2 we find three sources at 11.9 $\mu$m in the restored 
field of view (Fig.~\ref{timmi2}). We detect IRS1 itself as 
well as the most luminous member of the small star cluster 
$\approx 47''$ southeast of IRS1
\citep{Schreyer97}.
Since both sources have counterparts at 2.2 $\mu$m,
we can adopt the near--infrared astrometry for the restored TIMMI2 image. 
In doing so, we find that the third weak infrared source at 11.9 $\mu$m is 
located at the position of mm--source S1 (see Sect.~\ref{3mm}). 
Since we used the default chopping
throw of $20''$ in north--south direction, the S1 counterpart is unfortunately
placed very near to IRS1 in the chopped and nodded raw image. Although we could 
recover its true position by applying our image restoration algorithm,
the photometric information is strongly affected, so we will not report a
photometry here. 
At 4.6 $\mu$m we did not detect the S1 counterpart which speaks for a
deeply embedded object. However, we clearly detect the NIR object 8 also 
at 4.6 $\mu$m which further increases the reliability of our thermal infrared
astrometry. 
But more important, this detection reveals a clear
infrared excess for object 8, which provides strong evidence for dusty 
circumstellar (or circumbinary) material very near the related young 
stellar object(s).

%
%
             \subsection{CS $J$ = 2\pf1 data} 
%

\begin{figure*}[t] 
\vspace{16cm} 
\includegraphics{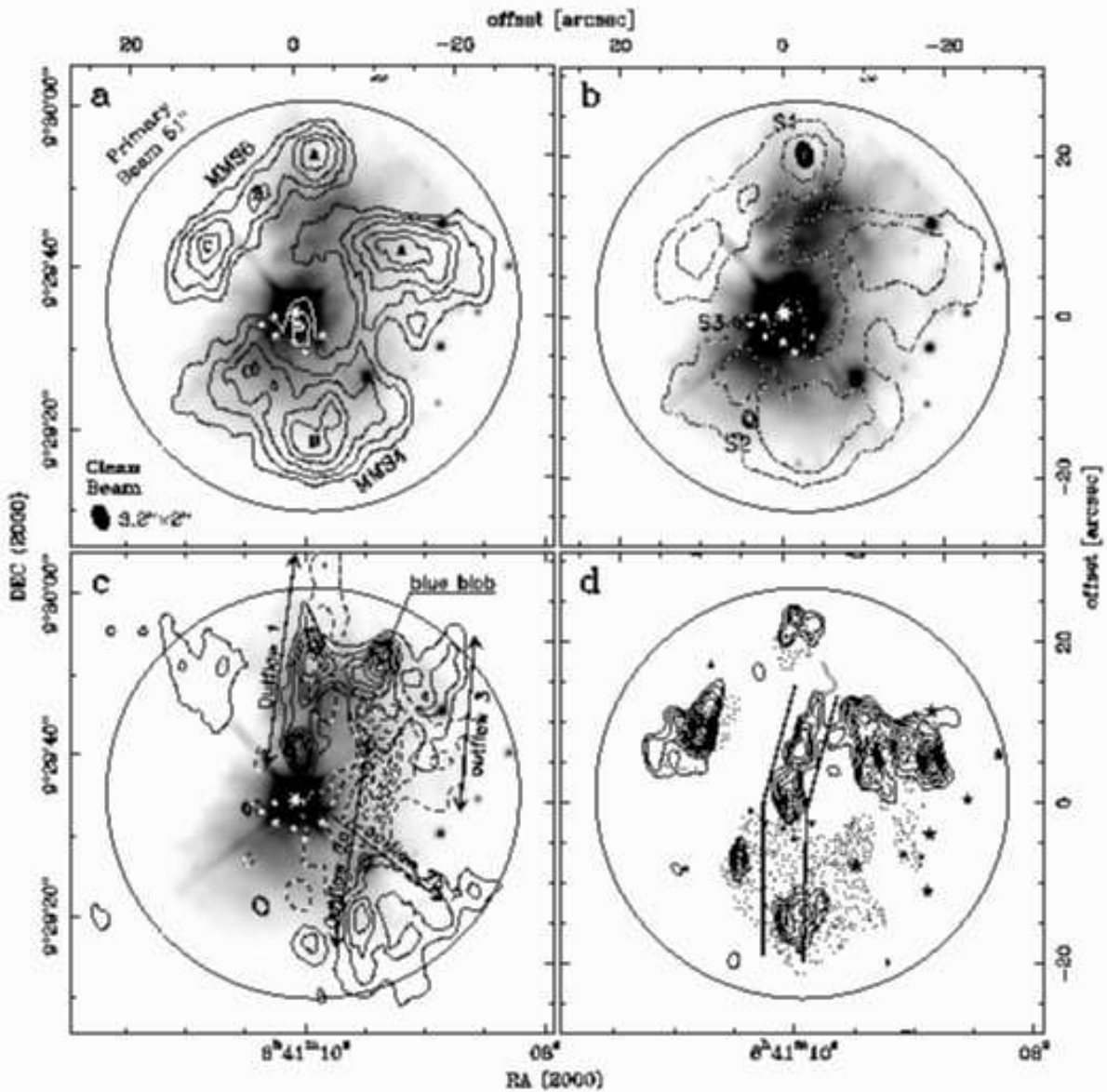} 
\figcaption{\small
        The PdBI data overlaid with the NTT $K$--band image, whereby the 
        image is shown with different lower cut levels to emphasise 
        different cloud structures of the background. 
        The diameter of the primary beam is always indicated with a large 
        circle. 
{\bf a} The total integrated CS    2$\rightarrow$1 line emission map   
        (continuum subtracted and zero--spacing corrected, 
        levels: 50\% to 90\% of the emission peak). 
        The size of the clean beam is shown in the lower left corner.
{\bf b} The three 3-mm continuum sources 
        (levels: 50\% to 90\% of the peak) are shown with solid contours.         
        The dotted lines indicate the 50\% and 70\% contours of the total 
        integrated CS line emission of Fig.\ a.
{\bf c} Integrated CS    2\pf1 line wing emissions 
        (red--shifted: 10.8$\le$\vlsr$\le$20 \kms$\!\!\!,$ dashed contours;
        blue--shifted: --3.4$\le$\vlsr$\le$5.2 \kms$\!\!\!,$ solid contours) 
        and the three continuum sources (thick ellipses = 50\% contour) 
        are superimposed.
{\bf d} Two channel maps of CS    2\pf1  (dashed: 
        \vlsr=8.6 \kms$\!\!\!,$ solid: 
        \vlsr=7.2 \kms$\!\!\!,$ 
        contour levels in percent of the peak values).  
        The positions of the stars and the gas stream structure 
        of the $K$--band image are indicated.  
\label{csfig}   
}  
\end{figure*}

Fig.\ \ref{csfig} shows the results of the PdBI measurements.
The total integrated CS    2\pf1 line emission   
(continuum subtracted and zero--spacing corrected, 
contour levels $\ge$50\% of the
peak emission, Fig.\ \ref{csfig}a) is overlaid 
with the NTT image for a better 
comparsion of the $K$--band objects.
CS emission with more than 30\% of the peak fills the entire area 
of the primary beam. 
The CS map implies that IRS1 is located in a low--density 
cavity surrounded by denser 
gas clumps which are possibly cut by the extent of the primary beam.
This morphology is in a good agreement with the H$^{13}$CO 
1\pf0 data obtained by Nakano et al.\ (2003).
If we assume that IRS1 is located in the cavity centre 
then the mean diameter of the cavity is $\approx 25''$ 
($\widehat{=} \; 2 \times 10^4$ AU) which 
matches the extent of the $K$--band nebulosity. Thus,
some of the dark cloud clumps are partly located in front or behind 
the cavity. However,  the turbulence inside the  
clouds is too large in order to find hints from their spectra 
about the true spatial locations.
Both the north- and the southeastern clumps 
(MMS\,4\,B/C and MMS\,5\,A/B/C see Fig.\ \ref{csfig}a) 
coincide with the more extended continuum emission peaks MMS 4 and 5 detected by 
Ward--Thompson et al.\ (2000) 
and have counterparts in H$^{13}$CO (Nakano et al.\ 2003: MMS5 = HC1 + HC2, MMS4 =
HC3 + HC4). 
However, the CS clump MMS 4A shows no well--defined counterpart
neither in the data of Ward--Thompson et al.\  nor in the data of 
Nakano et al. 
At the position of IRS1, no dense gaseous disk was found.  Only
a small gas clump (MMS 4D) was detected between the positions of IRS1 
and the low--mass   `object 3' which could be the
place of the origin of the twisted gas stream.
However, the uncertainty of the overlay positions of $\approx$2$''$  
makes it impossible to clarify the true association.

\begin{figure}
\vspace{4cm} 
\includegraphics{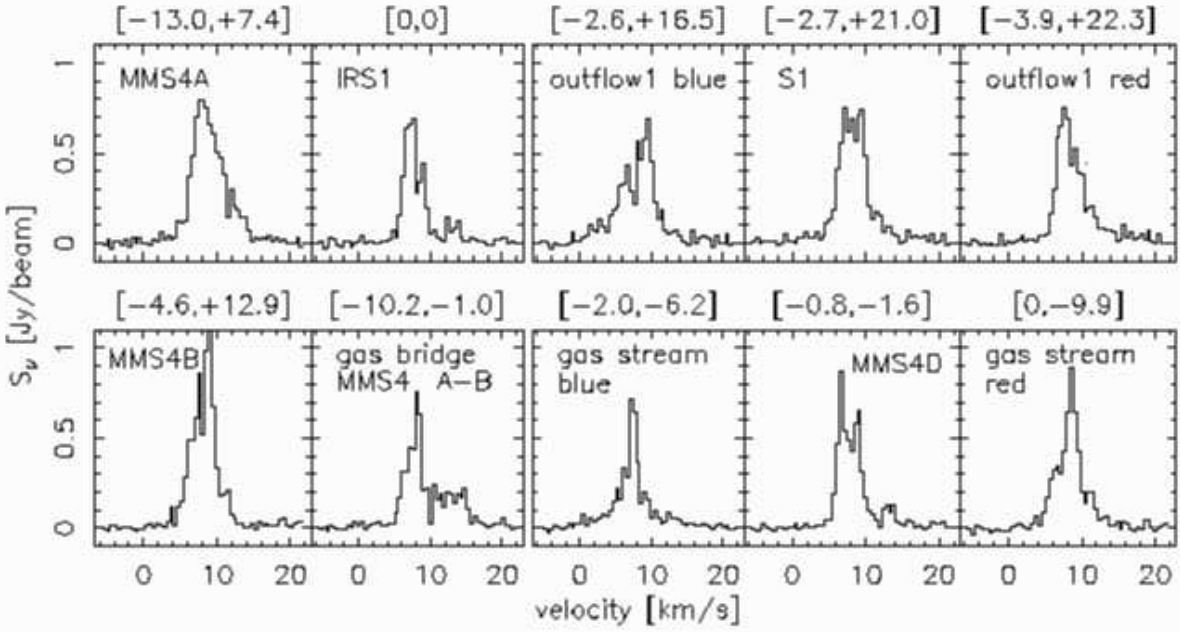} 
\figcaption{\small
CS 2\pf 1 spectra taken from different map positions.  
{The corresponding offset position in arcseconds
is given in brackets at the top of each spectrum.}
\label{sp}   
}  
\end{figure}

The study of the entire data cube implies that the dynamics 
of the gas in the 
observed sky region is very complex. In Sect.~\ref{s_out}, we will 
report different outflow systems and gas streams in much more detail. 
Fig.\ \ref{sp} displays selected spectra of different 
map positions. 
It is not clear, however, if the 
CS line is subsequently absorbed, or the spectra present a number 
of optically thin velocity components. The shape of the spectra 
varies strongly from position to position. In the observed area, 
there are hints that the emission at the peak positions of the 
denser cloud clumps can be absorbed (Fig.\ \ref{sp}: MMS 4A, IRS1, S1). 
However, at other selected positions,
e.g.\ the gas of the twisted $K$--band feature (see Sect.~\ref{s_out}, 
map in Fig.~\ref{csfig}d, and spectra in Fig.~\ref{sp}: gas stream blue, red),
the emission appears to be optically thin.

We estimated the gas mass of the dense cloud fragments inside the 
50\% contour level applying the formula (2) given by Nakamura et al.\ 
(1991). Although the assumptions of LTE and an optically
thin CS line emission are not met in our case, we can obtain a lower
mass limit using a kinetic temperature of 
20 K and a CS abundance of 
10$^{-9}\!\!.$ 
Table \ref{xx} summarises the integrated line emissions 
and the gas masses for several cloud clumps. 
Although we cover slightly differnet sky areas for the different cloud clumps, 
the mass estimates agree well with the results found by Nakano et al.\ (2003)
for the more extended continuum emission as well as for the H$^{13}$CO$^+$ 
measurements.

%
%
             \subsection{Continuum Sources \label{3mm} } 
%

Three rather weak 3-mm continuum 
point sources above the 3$\sigma$ level 
(S1, S2, and S3, see  Fig.~\ref{csfig}b) were detected. 
%
The coordinates of the point sources are listed in Table~\ref{pos} 
and agree well with the peaks of the more extended continumm emission 
regions MC1 (peak S1), MMS4 (peak S2), MMS5 (peak S3) by Nakano et al.\ (2003).
%
The PdB interferometer did not trace the more extended mm--emission seen by 
Ward--Thompson et al.\ (2000)
%
and Nakano et al.\ (2003).
%
The total integrated continuum fluxes are 
8.2 mJy, 4.0 mJy, and 1.4 mJy, and the peak fluxes amount to  
9.5,  6.4, and 5.1 mJy beam$^{-1}\!\!,$ respectively.  
Neither of these sources is coincident 
with IRS1.
The strongest mm--source S1 is located 20$''$ to the north. 
If we assume that the overlay positions are correct 
then the weakest mm--source S3 
is associated with the faint $K$--band object 7 in the NTT image.

Using a standard mass derivation algorithm as described by 
Henning et al.\ (2000),
we estimated gas masses of 3.5 \Mo\ for S1, 1.8 \Mo\ for S2, 
and 0.6 \Mo\ for S3,
using a dust temperature of 20 K, a distance of 800 pc, 
a gas--to--dust mass 
ratio of 150, and a mass absorption coefficient of 
$\kappa^{\rm d}_{\rm m}$ = 0.2 cm$^2$ g$^{-1}$
(Ossenkopf \& Henning 1994: thin icy mantles, gas density of 10$^6$ \cmmm),
and a density gradient $\propto r^{-2}$.

%
%
       \subsection{Outflows and gas streams \label{s_out}} 
%

The CS lines show subsequently rather strong 
and broad line wings. Fig.~\ref{csfig}c displays the results of the mapped 
red-- and blue--shifted line wings which shows a more complicated structure.
The  highly collimated bipolar outflow (outflow 1) to the north of IRS1 
whose probable driving source S1 we have detected in the 3-mm continuum as
well as at 11.9 $\mu$m is very remarkable. The blue outflow lobe seems to  
be stopped or redirected into our line of sight very close to IRS1. 
The outflow masses of the blue (and the mapped part of the red)
flow were estimated using the integrated CS intensities $\ge$ 3\,$\sigma$
which corresponds to the 30\% level contour of the map peaks. The blue 
flow contains 4.6\,10$^{-3}$ \Mo\ and the mapped red part 2.5\,10$^{-3}$ 
\Mo. For comparison, we estimate the masses traced with the CS 5\pf 4 
JCMT measurements (Schreyer et al.\ 1997): both the blue and red components 
contain 0.2 \Mo, indicating that the interferometer data trace only the 
densest flow parts. The estimation of the dynamical flow age is in this 
case not possible, since the blue flow is redirected and the red flow 
is cut by the primary beam.

Two bipolar outflows 2a and 2b are produced by the binary object 8 
with an angle between them 
of $\approx$62$^\circ$ in the sky plane.  
The outflow 2a is, however, much more 
intense
than the outflow 2b. The peaks of the red-- and 
blue--shifted CS emission of outflow 2a are roughly 8--10$''$ (projected 
distance = 6--8\,10$^3$AU) away from the central object. Since the red 
lobe of the outflow 2b is rather short and ceases near IRS1, we speculate 
that the binary object 8 can be located in front of IRS1. 
A third outflow (outflow 3) is associated with the infrared object 9.
Based on the number of outflows around IRS1, we can conclude that in 
the whole observed region around IRS1, the star--forming process is 
still on--going. With our images, we detect the {\em true} baby stars which 
are {\em not} the objects in the HST image presented by Thompson et al (1998).

The morphology of these outflows implies that IRS1 is not the origin of the 
bipolar outflow always detected with previous single--dish observations.

Fig.\ \ref{csfig}d shows two CS channel maps. One of the channel maps 
shows that the twisted jet--like $K$--band feature arising close to 
IRS1 has a counterpart in the CS gas in a very narrow velocity range 
between 7.34 and 7.89 \kms\ which fits very well the ``zig--zag'' 
structure of the gas stream. An opposite jet--like feature may be 
present in the second channel map. However, this gas pours into a 
denser cloud region. Thus the red--shifted gas stream is not so well 
separated as the blue--shifted one. Both channel maps clarify that 
the gas streams present undisturbed gas flows. The twisted shape may 
be produced by a precession of the central star. Based on the position 
uncertainties of the overlay, it is not really clear if IRS1 or the 
infrared object 3 is the origin of the bipolar gas stream or if there 
is a third obscured object between both sources. 
In addition, Fig.\ \ref{csfig}c  shows a blue--shifted gas blob at the 
``end'' of the infrared gas stream which seems to be produced by the 
piercing of the covering dark cloud.

%
%
             \section{Conclusions} 
%
%
%

We can conclude that IRS1 is in a more evolved evolutionary state   
than
the young B--type objects AFGL 490 
\citep{Schreyer02}
or G192.16--3.82 
(Shepherd, Claussen, \& Kurtz 2001) 
which show strong evidence for disks. 
These results showed that massive accretion 
disks orbiting B2--3 stars in the first 10$^3$--10$^5$ yrs of their 
main--sequence lifetime exist. However, these disks are unstable and 
might disappear, for instance, 
due to gravitational instabilities 
(e.g., Schreyer et al. 2002).
IRS1 seems to be in this phase where the 
accretion disk already disappeared and a small cavity was created, 
although, the star is still 
embedded in the centre of a more extended cloud core 
\citep{Krugel,Schreyer97}. 
The power of IRS1 leads to induced star formation 
in the surrounding denser cloud clumps, where 
a number of young stellar objects are 
embedded powering bipolar outflows. 
Our data show that the main source of the large-scale bipolar outflow 
is a deeply embedded young stellar object 20$''$ to the north of IRS1.
In addition, the object 8 in the closer environment of IRS1 
is a binary surrounded by  dusty circumbinary material 
and powering two bipolar outflows.

%
%
                        \begin{acknowledgements}                             
%

We acknowledge the help of the IRAM staff both of 
the Plateau de Bure and Grenoble. We  
especially thank Helmut Wiesemeyer for 
help provided during   data reduction.
The project was partly supported by DFG 
grants  He 1935/14-1.
\end{acknowledgements}


\bibliographystyle{apj}


\clearpage

\begin{deluxetable}{lcc}    
\tablecaption{ Positions of the new detected sources. The position of IRS1 (NICMOS) 
corresponds to Thompson et al. (1999). \label{pos} }  
  \tablehead{
    \colhead{object}&
    \colhead{RA($J$2000)} & 
    \colhead{DEC($J$2000)} \\
    \colhead{}&
    \colhead{(\ \ $^{\rm h}$ \ : \ $^{\rm m}$ \ : \ $^{\rm s}$)} & 
    \colhead{(\ \ $^\circ$ \ : \ $'$ \ : \ $''$)}  \\   
    \colhead{}&
    \colhead{[$\pm$0.1$^{\rm s}$]} & 
    \colhead{[$\pm$0.5$''$]}
    }    
\tablewidth{0pt}  
\startdata
IRS1 (NICMOS)    & 06:41:10.1  &    09:29:34.0\\
object 7         & 06:41:10.5  &    09:29:33.2 \\
S1               & 06:41:09.9  &    09:29:53.8  \\
S2               & 06:41:10.5  &    09:29:33.0 \\
S3               & 06:41:10.4  &    09:29:20.9  \\
C1               & 06:41:12.5  &    09:29:03.9 \\ 
\enddata
\end{deluxetable}


\clearpage
\begin{deluxetable}{lrrrrr}    
\tablecaption{Integrated CS fluxes and 
mass estimates for different cloud clumps labeled in 
Fig.~\ref{csfig}a. \label{xx} }  
  \tablehead{
    \colhead{cloud clumps}&
    \colhead{MMS\,4A} & 
    \colhead{MMS\,5A} & 
    \colhead{MM\,4B/C/D} & 
    \colhead{MMS\,5A/B/C} } 
\tablewidth{0pt}  
\startdata
$\int$$F$dv [Jy km\,s$^{-1}$] & 127.33& 32.16 & 184.25 &  94.77 \\
mass [\Mo ]                & 18.0  &   4.5 & 26.0   & 13.4 \\
  \enddata
\end{deluxetable}

\clearpage


\clearpage


\clearpage


\clearpage


\clearpage

\clearpage
\end{document}